\documentclass[aps,prd,amsmath,amssymb,
twocolumn,
floatfix
]{revtex4-1}
\usepackage{graphicx}
\usepackage{times}
\usepackage{multirow}
\usepackage{verbatim}
\usepackage{color,soul}
\usepackage{xcolor}
\usepackage{cancel}
\usepackage[normalem]{ulem}

\graphicspath{{figures/}}

\newcommand{\nuebar}{\ensuremath{\overline{\nu}_{e}} }
\newcommand{\uFive}{$^{235}$U}
\newcommand{\uEight}{$^{238}$U}
\newcommand{\pNine}{$^{239}$Pu}
\newcommand{\pOne}{$^{241}$Pu}
\newcommand{\xSec}{\ensuremath{\times10^{-43}~\text{cm}^2 / \text{fission} }}

\begin{document}

\title{Prospects for Improved Understanding of Isotopic Reactor Antineutrino Fluxes}

\author{Y. Gebre}%

\author{B. R. Littlejohn}
 \email{blittlej@iit.edu}

\author{P. T. Surukuchi}%
 \email{psurukuc@hawk.iit.edu}

\affiliation{Physics Department, Illinois Institute of Technology, Chicago, IL 60616, USA}%

\begin{abstract}
Predictions of antineutrino fluxes produced by fission isotopes in a nuclear reactor have recently received increased scrutiny due to observed differences in predicted and measured inverse beta decay (IBD) yields, referred to as the `reactor antineutrino flux anomaly.'  
In this paper, global fits are applied to existing IBD yield measurements to produce constraints on antineutrino production by individual plutonium and uranium fission isotopes.  
We find that fits including measurements from highly \uFive-enriched cores and fits including Daya Bay's new fuel evolution result produce discrepant best-fit IBD yields for \uFive~and \pNine.
This discrepancy can be alleviated in a global analysis of all datasets through simultaneous fitting of \pNine, \uFive, and \uEight~yields.  
The measured IBD yield of \uEight~in this analysis is (7.02 $\pm$ 1.65) \xSec, nearly two standard deviations below existing predictions.  
Future hypothetical IBD yield measurements by short-baseline reactor experiments are examined to determine their possible impact on global understanding of isotopic IBD yields.  
It is found that future improved short-baseline IBD yield measurements at both high-enriched and low-enriched cores can significantly improve constraints for \uFive, \uEight, and \pNine, providing comparable or superior precision to existing conversion- and summation-based antineutrino flux predictions.
Systematic and experimental requirements for these future measurements are also investigated.
\end{abstract}

\maketitle

\section{Introduction}\label{sec:introduction}

Nuclear reactors are prolific, pure sources of electron antineutrinos.  
As the primary parent isotopes \uFive, \uEight, \pNine, and \pOne~fission inside an operating nuclear reactor core, their neutron-rich fission products undergo successive beta decays, producing fluxes of antineutrinos with energies in the $\sim$0-8~MeV range.  
While these decay products are important to understand for the purposes of nuclear power generation, the antineutrino flux from these products has also received attention in particle physics due to its usefulness in probing the nature of the weak interaction and fundamental properties of neutrinos.  
Reactor antineutrino flux measurements have been used in the past to verify predictions of the charged~\cite{bib:reinesCC,bib:B4, bib:rovnoCC} and neutral current~\cite{bib:reinesNC,bib:Krasno3, bib:TEXONO} weak interaction.  
Over the past two decades, reactor antineutrinos have been utilized to produce measurements of Standard Model neutrino oscillation parameters~\cite{KamLAND,bib:prl_rate,bib:reno,dc_bump, bib:prd_osc}.  

A variety of theoretical methods have historically been used to calculate the expected energy spectrum and integrated flux of betas and antineutrinos produced in nuclear reactors.  
Early efforts utilized statistical and theoretical assumptions based on tenets of nuclear physics~\cite{Wigner}.  
Following sustained investigation of fission product yields and beta decay schemes and organization of standard nuclear data tables, antineutrino spectra could be calculated via the `summation' (or \textit{ab initio}) approach~\cite{Greenwood,Klapdor}.  
In the 1980s, high-precision spectroscopy of fission product betas~\cite{bib:ILL_1, bib:ILL_2, bib:ILL_3} enabled a more precise calculation of the antineutrino spectra from \uFive, \pNine, and \pOne~via the beta conversion method~\cite{Vogel}.  
The latter approach produced good agreement with existing reactor antineutrino flux and spectrum measurements at that time~\cite{bib:Bugey3}.  

More recently, re-calculations of beta-converted reactor antineutrino spectra have produced changes in predicted fluxes and spectra~\cite{bib:mueller2011,bib:huber}, resulting in new disagreements with respect to measured reactor antineutrino fluxes~\cite{bib:mention2011}.  
Flux measurements by the Daya Bay experiment have subsequently validated the existence of this flux discrepancy~\cite{bib:prl_reactor}.  
Further, results by the $\theta_{13}$ reactor experiments have uncovered a discrepancy in the measured antineutrino spectrum with respect to beta-converted predictions, with particular focus on the 5-7~MeV region of antineutrino energy~\cite{bib:prl_reactor,reno_bump,dc_bump}.  

The cause of these anomalies has been a source of speculation within the nuclear and particle physics communities.  
The measured antineutrino flux deficit could be explained via the existence of oscillations to eV-scale mass sterile neutrinos~\cite{bib:mention2011,bib:kopp,bib:giuntiGlobal}.  
A variety of explanations related to antineutrino production in the core have also been provided to explain both the spectrum and flux anomalies.  
Some studies have pointed out possible explanations stemming from the beta-converted prediction, such as incorrect beta spectrum measurements~\cite{bib:dwyer,hayes2} or incorrect aspects of the beta-to-$\overline{\nu}_e$ energy conversion procedure~\cite{bib:hayes, bib:hayesShape}.  
Other studies have renewed the use of summation predictions as a tool for understanding the anomalies' possible origins~\cite{bib:fallot,bib:dwyer,bib:sonzogni,sonzongi2,hayes2}.  
Some of these studies have introduced the hypothesis that one of the four primary fission isotopes may be primarily culpable for one or both anomalies~\cite{hayes2,haser}.  

Given the scope of possible implications, recent effort has been made to develop antineutrino-based analyses and datasets to provide additional insights into the flux and spectrum anomalies. 
In the coming years, a new generation of short-baseline experiments based at highly-enriched uranium reactors will obtain new datasets to directly probe sterile neutrino oscillations and measure \nuebar production by \uFive~\cite{prospect,stereo,solid}.  
A number of recent analyses have attempted to exploit differences in reactor fuel content between datasets: by comparing measurements of differing mixtures of \uFive, \uEight, \pNine, and \pOne ~\nuebar fluxes at low-enriched (LEU) reactors or these to measurements of pure $^{235}$U-produced fluxes at highly-enriched (HEU) reactors, these analyses have probed \nuebar production rates by individual fission isotopes.  
Measurement of variation in inverse beta decay (IBD) detection rates arising from evolution of reactor fuel content at the Daya Bay experiment has provided evidence that incorrect modeling of \nuebar production by individual fission isotopes must be at least partially responsible for the flux anomaly~\cite{bib:prl_evol}. 
New global fits comparing measured IBD spectra~\cite{Huber:2016xis} and integrated yields~\cite{Giunti,Giunti2} between reactor experiments have suggested that modeling of \uFive~\nuebar production, in particular, may be incorrect.  
Recent inclusion of the Daya Bay evolution dataset in these fits has produced mixed results, with a preference for `composite models' including both incorrect flux predictions and sterile neutrino oscillations~\cite{GiuntiMe}.  

In this study, we utilize a IBD yield measurement global fit to further explore current and possible future constraints on \uFive, \uEight, and \pNine~\nuebar production in reactors.  
We begin by reproducing aspects of the global fit in~\cite{bib:prl_evol,Giunti2,GiuntiMe} as a consistency check, while highlighting discontinuities present between measurements at Daya Bay and at HEU cores.  
We then show that a combined fit to all IBD yield data can also be utilized to produce a measurement of \nuebar production by \uEight, and that the central value of this measurement differs from existing predictions.  
By replacing existing measurements in our global fit with hypothetical future precision measurements of HEU and LEU IBD yields, we then investigate future achievable levels of precision for \uFive, \uEight, and \pNine.
We find that future IBD yield measurements have the capability to produce comparable or superior bounds on all three isotopes relative to existing conversion- or summation-based flux predictions. 
We then investigate the systematic and experimental requirements for achieving this improved understanding in future measurements.  

We begin in Section~\ref{sec:global} with a description of the global fit.  
Results of the fit and constraints on \uEight~are presented and discussed in Sections~\ref{subsec:constrain}, ~\ref{sec:model_comps}, and~\ref{subsec:unconstrain}.   
In Section~\ref{sec:SBL}, we describe the set of considered future hypothetical experiments and the result of applying global fits to the hypothetical results of these experiments.  
Experimental and systematic requirements for future experiments are discussed in Section~\ref{subsec:Syst}.
Main results are then summarized in Section~\ref{sec:summary}.

\section{Global Fits to Existing Reactor IBD Yield Measurements}
\label{sec:global}

Due to differing levels of antineutrino production per fission for the primary fission isotopes, the total number of inverse beta decays (IBDs) detected by a reactor antineutrino experiment depends on the isotopic content of nearby reactors.  In the case of a single nearby reactor, the number of detected IBDs in a time interval $dt$ can be described as:
\begin{equation}\label{equ_ibd_rate}
\frac{dN}{dt} =  \frac{ N_p  \varepsilon
\!P(L)}{4\pi L^2} \frac{W_{th}(t) \sigma_f(t)}{\bar{E}(t)},
\end{equation}
where
$N_p$ is the number of target protons,
$\varepsilon$ is the efficiency of detecting IBDs, 
$P(L)$ is the survival probability due to neutrino
oscillations,
$L$ is the core-detector distance,
and $W_{\mathrm{th}}(t)$ is the reactor's thermal power.  
The term $\bar{E}(t) = \sum_{i}F_{i}(t)e_{i}$ is the core's average energy released per fission, where 
$F_i(t)$ are the fission fraction of fission isotope $i$ and
 $e_i$ is the average energy released per fission of isotope $i$.
The term $\sigma_f(t)$ is the IBD cross-section per fission, or IBD yield:
\begin{equation}\label{eq:Iso1}
\sigma_f(t) = \sum_i F_i(t) \sigma_i,
\end{equation}
where $\sigma_i$ is the IBD yield per fission for each parent fission isotope.  

Measurement of IBD detection rates $dN/dt$ has enabled a variety of reactor antineutrino experiments to generate one or multiple IBD yield values $\sigma_f$ spanning a wide range of fission fractions.  
Some experiments have utilized highly-enriched \uFive-burning reactors to measure $\sigma_f$ for a single set of fission fractions~\cite{bib:Krasno1,bib:Krasno2,bib:Krasno3,bib:ILL_nu,bib:nucifer,bib:srp}.  
Other experiments reporting $\sigma_f$ measurements have utilized commercial low-enriched uranium reactors experiencing substantial fission fraction variations over the course of their fuel cycles.
Most of these datasets have provided one $\sigma_f$ measurement for a single set of fission fractions that corresponds to the average over the run period of the experiment~\cite{bib:gosgen,bib:Rovno1,bib:Rovno2,bib:B4,bib:Bugey3,bib:chooz}.  
Recently, the Daya Bay experiment separately analyzed data from time periods of differing fission fraction to produce eight $\sigma_f$ measurements covering a range of fission fraction values between 0.25 and 0.35~\cite{bib:prl_evol}.  

As demonstrated by Eq.~\ref{eq:Iso1}, the multiple fission fraction and $\sigma_f$ values can be utilized to determine IBD yields for the individual fission isotopes, $\sigma_i$.  
In this study, we refer to these isotopic IBD yields as $\sigma_5$, $\sigma_8$, $\sigma_9$, $\sigma_1$ for the isotopes \uFive, \uEight, \pNine, \pOne, respectively.
To obtain isotopic IBD yields in this analysis, we use a least-squares test statistic similar to those utilized in previous studies~\cite{Giunti, Giunti2, bib:prl_evol}:

\begin{equation}
\begin{gathered}
\label{eq:Iso3}
\chi^2 = \sum_{a,b}\bigg(\sigma_{f,a} - r \sum_i F_{i,a} \sigma_i\bigg)
	\textrm{V}^{-1}_{ab}
    \bigg(\sigma_{f,b} - r \sum_i F_{i,b} \sigma_i\bigg) \\
     + \sum_{i,j}(\sigma^{th}_{i}-\sigma_{i}) \textrm{V}^{-1}_{\textrm{ext},ij}(\sigma^{th}_{j}-\sigma_{j}).
\end{gathered}
\end{equation}

In this fit, experimental inputs are $F_i$ and $\sigma_f$ as described above, as well as the covariance matrix $V_{ab}$ describing uncertainties of the measurements $\sigma_f$.  
For all experiments excepting Daya Bay, this study utilizes $\sigma_f$ and $F_i$ values from Ref.~\cite{Giunti}, which attempts to reconcile differences in theoretical inputs between the different measurements.  
The covariance matrix $V_{ab}$ for the Daya Bay evolution data is taken from Ref.~\cite{bib:prl_evol}, while for the remaining measurements, $V_{ab}$ is once again obtained from Ref.~\cite{Giunti}.  

The primary fit parameters are the four isotopic IBD yields, $\sigma_i$.  
To enable meaningful constraints on the dominant fission isotopes \uFive~and \pNine, or to test  hypotheses regarding the origin of the reactor antineutrino flux anomaly, previous fits have applied a variety of external constraints on $\sigma_i$ values.  
The second term in Eq.~\ref{eq:Iso3} enables this by applying a penalty as  isotopes deviate from their theoretically-predicted values $\sigma_i^{th}$, which are taken from Ref.~\cite{bib:huber} for $\sigma^{th}_{5,9,1}$ and Ref.~\cite{bib:mueller2011} for $\sigma^{th}_{8}$.  
The form of the covariance matrix V$_{\textrm{ext}}$ will vary depending on the fit scenario, and will be explicitly described as each fit is introduced and applied.  
Additionally, a scaling factor $r$ can be fitted to allow common rescaling of theoretical IBD yields if desired; when not being fitted, $r$ is set to unity.


\subsection{Isotopic IBD Yield Fits Utilizing $\sigma_{8,1}$ Constraints}\label{sec:global_fits}
\label{subsec:constrain}

To begin, we utilize existing IBD yield datasets to determine best-fit yields for the dominant fission isotopes \uFive~and \pNine, examining discontinuities between the different datasets in the process.  
In a conventional reactor core, the fission fractions of the two plutonium isotopes both rise with burnup.  
Thus, the fit described by Eq.~\ref{eq:Iso3} will contain a major degeneracy between the parameters $\sigma_{9}$ and $\sigma_{1}$. 
In addition, due to the small values of $F_{238}$ in the existing datasets, the fit will provide a comparatively weaker constraint on $\sigma_8$.  
To address these issues and enable meaningful constraints on $\sigma_{5}$ and $\sigma_{9}$, $\sigma_8$ and $\sigma_1$ are treated as nuisance parameters in the fit:
\begin{equation}
\begin{gathered}
\label{eq:Iso4}
\chi^2 = \sum_{a,b}\bigg(\sigma_{f,a} - \sum_i F_{i,a} \sigma_i\bigg)
	\textrm{V}^{-1}_{ab}
    \bigg(\sigma_{f,b} - \sum_i F_{i,b} \sigma_i\bigg) \\
     + \sum_{i,j=238,241}(\sigma^{th}_{i}-\sigma_{i}) \textrm{V}^{-1}_{\textrm{ext},ij}(\sigma^{th}_{j}-\sigma_{j}).
\end{gathered}
\end{equation}
To minimize the dependence of the fit on the theoretical inputs, the terms V$_{\textrm{ext},8,8}$ and V$_{\textrm{ext},1,1}$ of the diagonal 2x2 matrix V$_{\textrm{ext}}$ are set to 10\% of the theoretical central values, respectively.
This expression is effectively identical to the approach taken in the Daya Bay fuel evolution analysis~\cite{bib:prl_evol}.  

\begin{figure}[htb!pb]
\centering
\includegraphics[width=1.1\linewidth]{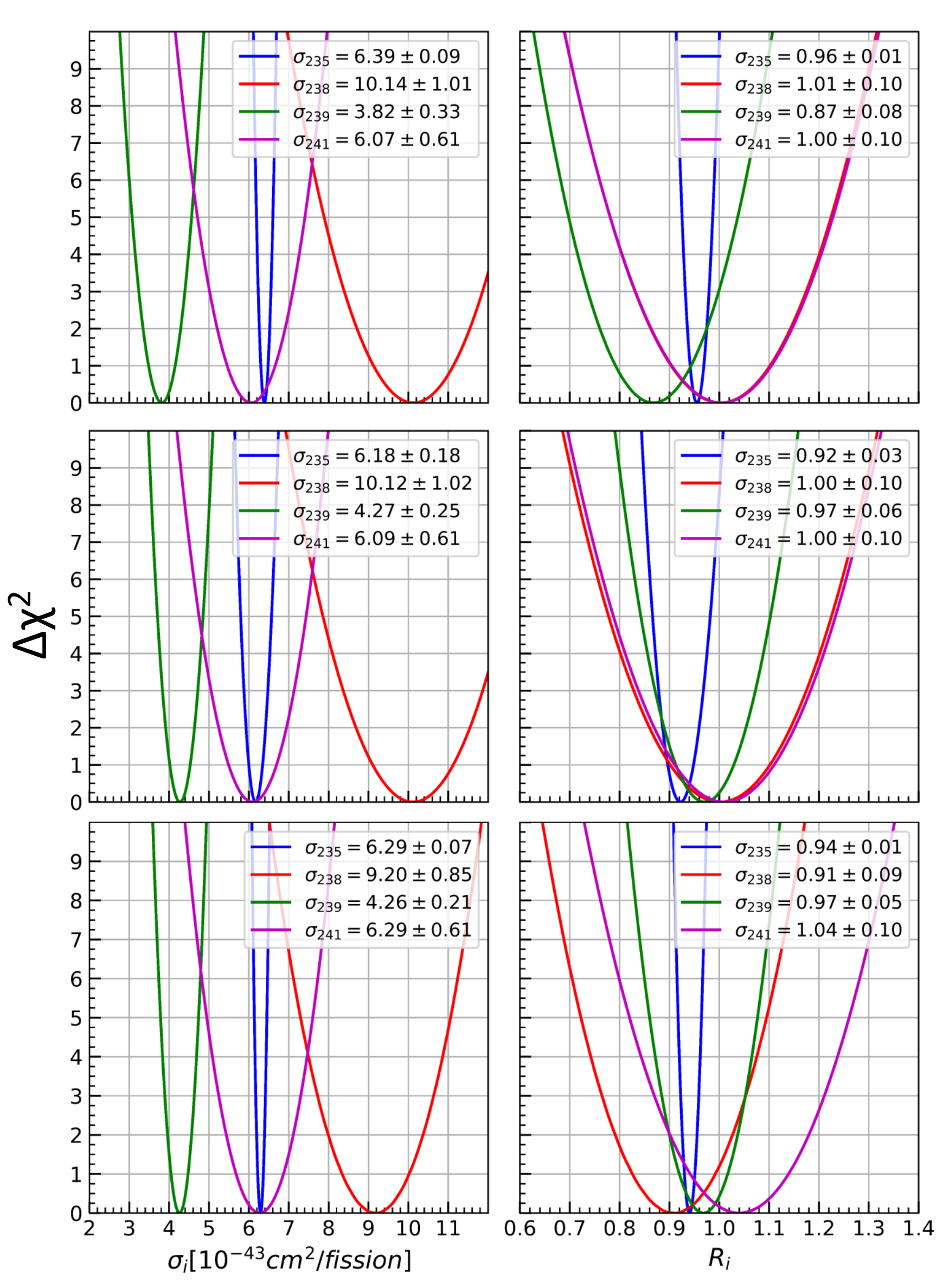}
\caption{One-dimensional $\chi^2$ profiles for the IBD yields $\sigma_{5,8,9,1}$ obtained by applying the fit of Eq.~\ref{eq:Iso4} to global rate data (top), Daya Bay evolution data (middle), and the combined dataset (bottom).  The fit includes external constraints on $\sigma_8$ and $\sigma_1$ as described in the text.  Profiles are provided in terms of \xSec~(left) and in terms of $R_i = \sigma^{meas}_{i}/\sigma^{th}_{i}$ (right), the ratio of predicted and measured IBD yields.}
\label{fig:Fit1}
\end{figure}

The result of the fit with constrained $\sigma_1$ and $\sigma_8$ is shown in Figure~\ref{fig:Fit1} for three datasets: all reactor rate measurements excluding the recent Daya Bay evolution result (referred to as the `global rate data'), the Daya Bay evolution result by itself, and the combination of both datasets.  
The fits are represented in this figure by one-dimensional $\chi^2$ profiles, where the profile for each isotope includes a full minimization over the other parameters.  The best-fit $\sigma_i$ value and 1$\sigma$ widths are also pictured.  
In addition, these results are pictured in terms of the ratio between measured and predicted IBD yields, $R_i = \sigma^{meas}_{i}$/$\sigma^{th}_{i}$.  
Minimum $\chi^2$ and $\chi_{min}^2$/NDF values for each of the three fits are also provided in Table~\ref{tab:hyp_nc}.

\begin{table}[hptb!]
\centering
\begin{tabular}{c|c|c|c|c|c}
\hline
\multirow{3}{*}{Dataset} & \multirow{3}{*}{$\chi^2_{min}$} & \multirow{3}{*}{$\frac{\chi^2_{min}}{\textrm{NDF}}$}& \multicolumn{3}{|c}{Hypothesis $\Delta \chi^2_{min}$} \\ \cline{4-6}
&& & Common & \uFive & \pNine \\
&& & Error & Only & Only  \\ \hline \hline
Global Rates &16.72& 0.70& 1.06 &3.32 &2.63  \\ \hline
Daya Bay &3.60 &0.60& 5.07 &0.25& 5.17  \\ \hline
Combined &23.49 &0.76& 3.78 & 0.90&6.66   \\ 
\hline
\end{tabular}
\caption{Summary of goodness-of-fit values for various datasets and fit implementations.  $\chi_{min}^2$ and $\chi^2_{min}$/NDF values are for the fit of Eq.~\ref{eq:Iso4} with nuisance parameters $\sigma_8$ and $\sigma_1$.  $\Delta\chi^2_{min}$ values are provided with respect to this $\chi^2_{min}$ for similar fits where alternate external fit constraints are applied in order to test various hypotheses regarding the origin of the reactor antineutrino flux anomaly, as described in the text.}
\label{tab:hyp_nc}
\end{table}


In the profiles shown in Figure~\ref{fig:Fit1}, differences can be seen in best-fit values between differing datasets.  
The best-fit $\sigma_5$ value from the global rate data, which is dominated by the HEU measurements of Refs.~\cite{bib:ILL_nu,bib:srp,bib:Krasno3}, appears at (96$\pm$1)\% of the Huber-predicted value, while for Daya Bay evolution data, this value is (92$\pm$3)\%.  
For $\sigma_9$, the Daya Bay evolution best-fit value of (97$\pm$6)\% of the Huber-predicted value once again differs  from the (87$\pm$8)\% value produced by a fit to the global rate data.  
These discontinuities are also reflected in the \textbf{235+239} global fit of Ref.~\cite{GiuntiMe}.  
This discontinuity between Daya Bay evolution and global rate datasets naturally results in a higher $\chi_{min}^2$/NDF in the combined fit (0.76 = 23.49/31), than for either the Daya Bay evolution (0.60 = 3.60/6) or global rate (0.70 =16.72/24) fits separately.  
The global rate measurement of $\sigma_5$ is overwhelmingly determined by the HEU dataset: a similar fit of the HEU data by itself yields a best-fit value of (6.39$\pm$0.10)~\xSec, nearly identical to the fit of all global rate data shown in Figure~\ref{fig:Fit1}.  
Thus, it appears that the fundamental underlying discontinuity in the combined fit of all data is between the Daya Bay evolution and HEU datasets.  
Indeed, if HEU measurements are removed from the combined dataset, the $\chi^2$/NDF is reduced to 0.49 = 11.8/24.  

\subsection{Comparing Reactor Flux Anomaly Hypotheses}
\label{sec:model_comps}

The discontinuity between datasets can also be examined by statistically comparing the best fits obtained above to a range of hypotheses regarding the origin of the reactor antineutrino flux anomaly. 
In this study, we choose to examine only scenarios related to improper modeling of neutrino production using the conversion approach.  
In the near future, scenarios involving sterile neutrino oscillations will be rigorously tested without the use of IBD yield measurements, via searches for reactor antineutrino spectral distortion over multiple baselines.  
Following these $L/E$-based searches, global IBD yield data can be expressly utilized for probing errors in beta-converted flux modelling.  

Conversion-produced flux predictions may only be improperly predicting one of the two dominant fission isotopes, \uFive~or \pNine.  
Such a scenario could be produced by isolated flaws in the beta spectrum measurement of a single fission isotope, or by systematic mis-modeling of the yields~\cite{hayes2} or beta decay shape corrections~\cite{bib:hayesShape,bib:bayes} of daughter isotopes from one of these fission isotopes.  
The compatibility of this scenario with the combined IBD yield data was represented with a fit model similar to the one described above.  As an example, the model for the `\uFive-only' hypothesis is
\begin{equation}
\begin{gathered}
\label{eq:Iso5}
\chi^2 = \sum_{a,b}\bigg(\sigma_{f,a} - \sum_i F_{i,a} \sigma_i\bigg)
	\textrm{V}^{-1}_{ab} 
    \bigg(\sigma_{f,b} - \sum_i F_{i,b} \sigma_i\bigg) \\
     + \sum_{i,j=8,9,1}(\sigma^{th}_{i}-\sigma_{i}) \textrm{V}^{-1}_{\textrm{ext},ij}(\sigma^{th}_{j}-\sigma_{j}).
\end{gathered}
\end{equation}
This model contains one unconstrained fit parameter $\sigma_{5}$, in addition to nuisance parameters for the remaining isotopes, constrained according to the existing theoretical predictions.  
For the `\uFive-only' and `\pNine-only' models, V$_{\textrm{ext}}$ is taken from Ref.~\cite{GiuntiGlobalNew}.

On the other hand, improper characterization of the BILL spectrometer used for \uFive, \pNine, and \pOne~beta spectrum measurements~\cite{bib:BILL}, or a common conversion mis-modeling effect could yield equally inaccurate conversion predictions for all three of these isotopes.  
This scenario, termed the `common conversion error' scenario, was fit to the data using the following model: 
\begin{equation}
\begin{gathered}
\label{eq:Iso5}
\chi^2 = \sum_{a,b}\bigg(\sigma_{f,a} - r \sum_{i=5,9,1} F_{i,a} - F_{8,a} \sigma_8\bigg)
	\textrm{V}^{-1}_{ab} \\
    \times\bigg(\sigma_{f,b} - r \sum_{i=5,9,1} F_{i,b} - F_{8,b} \sigma_8\bigg) \\
     + \sum_{i,j=5,8,9,1}(\sigma^{th}_{i}-\sigma_{i}) \textrm{V}^{-1}_{\textrm{ext},ij}(\sigma^{th}_{j}-\sigma_{j}).
\end{gathered}
\end{equation}
This `common conversion error' model contains one unconstrained fit parameter $r$ (a common scaling of the conversion-predicted yields $\sigma_{5,9,1}$) in addition to nuisance parameters for all isotopic IBD yields, constrained once again according to the existing theoretical predictions detailed in Ref.~\cite{GiuntiGlobalNew}.

The compatibility of each hypothesis with the lesser-constrained simultaneous fit of $\sigma_5$ and $\sigma_9$ can be judged by the $\Delta \chi^2_{min}$ between the two cases; $\Delta \chi^2_{min}$ values for the various scenarios and datasets are given in Table~\ref{tab:hyp_nc}.  
In general, both datasets (particularly the Daya Bay evolution data) tend to disfavor the hypothesis that \pNine~is the sole contributor to the reactor flux anomaly.  
However, Daya Bay evolution data strongly favors the \uFive-only hypothesis over the common conversion error hypothesis, while global rate data tends to favor the common conversion error hypothesis over the \uFive-only hypothesis.  
As a consequence of this discontinuity between HEU and Daya Bay evolution datasets, the combined fits do not reflect a strong preference for either of these two hypotheses.  

\subsection{Investigation of \uEight~Constraints}\label{sec:uEight}
\label{subsec:unconstrain}

It is interesting to note in Figure~\ref{fig:Fit1} that the combination of Daya Bay evolution and global rate datasets results in a substantial shift in $\sigma_8$, from 10.12 \xSec~and 10.14 \xSec~in the DYB-only and global rate cases, respectively, to 9.20 \xSec~in the combined case.   
This feature was further investigated by removing the external constraint on $\sigma_8$ entirely in Eq.~\ref{eq:Iso4}:
\begin{equation}
\begin{gathered}
\label{eq:Iso5}
\chi^2 = \sum_{a,b}\bigg(\sigma_{f,a} - \sum_i F_{i,a} \sigma_i\bigg)
	\textrm{V}^{-1}_{ab}
    \bigg(\sigma_{f,b} - \sum_i F_{i,b} \sigma_i\bigg) \\
    + \frac{(\sigma^{th}_{1}-\sigma_{1})^2}{V_{\textrm{ext},11}}.
\end{gathered}
\end{equation}

\begin{figure}[htb!pb]
\centering
\includegraphics[width=0.95\linewidth]{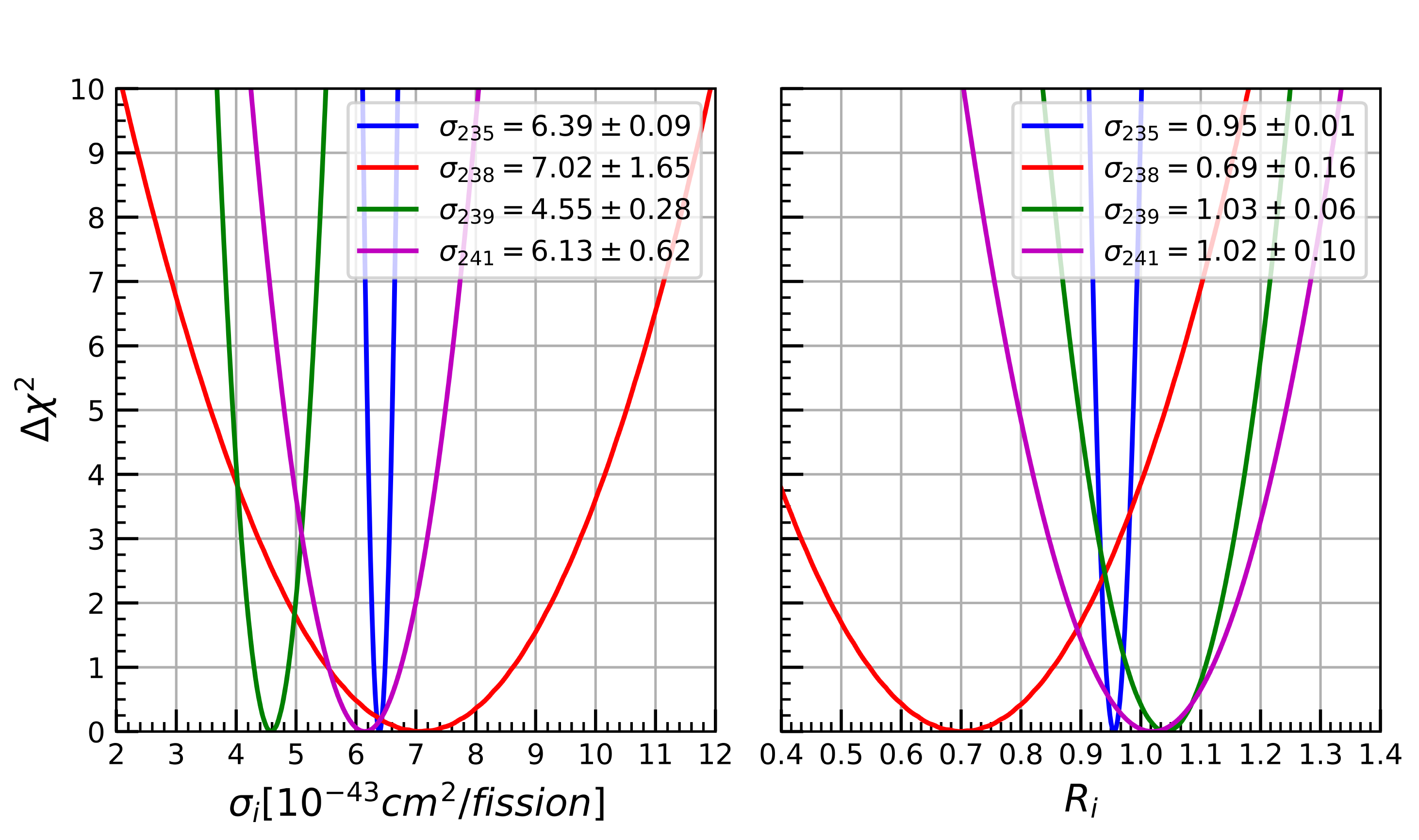}
\caption{Profile $\chi^{2}$ histograms for a combined fit of Daya Bay evolution data and the global rate data with no constraints on \uEight. Profiles are provided in terms of \xSec~(left) and in terms of $R_i$ (right), the ratio of predicted and measured IBD yields.}
\label{fig:U238_Det}
\end{figure}

The combined fit is shown in Figure~\ref{fig:U238_Det}.  The removal of the $\sigma_8$ constraint improves the $\chi^2_{min}$/NDF of the combined fit modestly, to 0.65 = 20.76/32.  
The fit produces non-negligible bounds on all three values $\sigma_5$, $\sigma_9$, and $\sigma_8$:
\begin{eqnarray}
\centering
\label{eq:Iso6}
\sigma_5 & = & (6.39 \pm 0.09) \xSec \\ 
\sigma_9 & = & (4.55 \pm 0.28) \xSec \\
\sigma_8 & = & (7.02 \pm 1.65) \xSec
\end{eqnarray}
The non-negligible constraint on $\sigma_8$ is enabled by the difference in \uEight-produced \nuebar contributions between the highly enriched ($\sim$0\%) and conventional ($\sim$10\%) reactor core measurements, combined with the HEU-independent constraints enabled by Daya Bay's evolution measurement.  
If a similar fit is applied excluding either the Daya Bay evolution data or all HEU data, no meaningful constraints can be obtained for $\sigma_8$, and $\sigma_5$ and $\sigma_9$ uncertainties are significantly inflated.  
This direct measurement of  $\sigma_8$ is almost two standard deviations below the predicted central value of 10.1\xSec.  
It is also worth noting that while this measured yield is higher than that of the other three primary fission isotopes, it deviates somewhat from the (3Z-A) systematic trend in IBD yields noted in Ref.~\cite{bib:sonzogni}.  

\begin{figure}[htb!pb]
\centering
\includegraphics[width=0.95\linewidth]{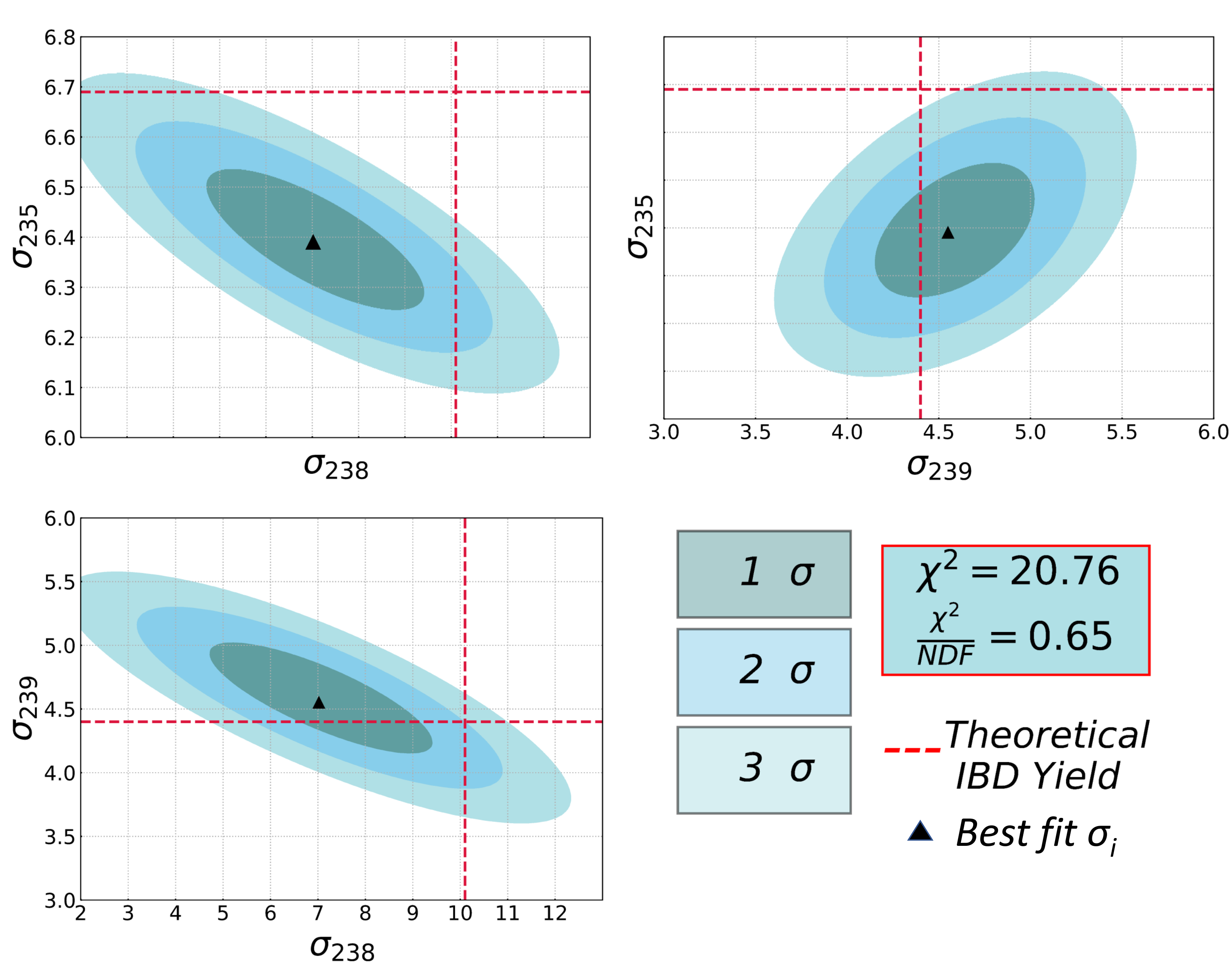}
\caption{Two-dimensional 1-, 2-, and 3-$\sigma$ allowed regions and predicted central values for the IBD yields $\sigma_{5,8,9}$ obtained by applying the fit of Eq.~\ref{eq:Iso5} to the combination of global rate and Daya Bay evolution datasets. 
IBD yields are provided in units of 10$^{-43}$cm$^2$/fission.  
Also shown in red dashed lines are the theoretical IBD yields.}
\label{fig:U238_Det_Triangle}
\end{figure}

Correlations between fitted $\sigma_{5,9,8}$ values are shown in Figure~\ref{fig:U238_Det_Triangle}.
The fitted $\sigma_8$ shows a significant degree of anti-correlation with both $\sigma_5$ and $\sigma_9$.
Thus, since removing external constraints on $\sigma_8$ results in a decrease in its best-fit value, the best-fit $\sigma_5$ is found at 96\% of the Huber-predicted value, slightly higher than the value obtained in the previous combined fit.  
Likewise, the best-fit $\sigma_9$ increases by roughly 1 standard deviation from the previously-discussed combined fit, to 103\% of the Huber-predicted value.  

After considering this result, a variety of scenarios appear capable of resolving discontinuities present in the global IBD yield picture: 
\begin{itemize}
\item{Composite models, which include both sterile neutrino oscillations and incorrect flux predictions for \uFive~or \pNine, are correct, as described in Ref.~\cite{GiuntiMe}.}
\item{\uEight~fluxes, as well as \uFive~and/or \pNine~fluxes, are incorrectly predicted.}
\item{Portions of the existing global dataset, specifically HEU IBD yield measurements or Daya Bay's evolution result, are incorrect in some way.}
\end{itemize}
Improved IBD yield data and short-baseline oscillation data from HEU and LEU reactors will be required to determine which of these hypotheses are correct.  


\section{Future Improvements From New Short-Baseline Measurements}
\label{sec:SBL}

We have demonstrated how existing IBD rate measurements can be used to constrain individual IBD yields $\sigma_i$ and examine hypotheses relating to the origin of the reactor flux anomaly.  
We will now investigate the extent to which future reactor experiments may improve $\sigma_i$ constraints and enhance the clarity of the global IBD yield picture. 

\subsection{Definition of Experimental Parameters}

A variety of new ton-scale compact reactor antineutrino experiments have either recently been deployed~\cite{bib:danss_2016,bib:neos,bib:neutrino4} or will be deployed in the near future~\cite{prospect,stereo,solid} at short baselines from operating LEU~\cite{bib:danss_2016,bib:neos} or HEU~\cite{prospect,stereo,solid} reactors.  
These experiments' detectors incorporate a wide variety of detection technologies and background rejection techniques to detect an IBD signal despite sizable cosmogenic and ambient backgrounds in their near-surface, near-reactor environments.  
These technology and analysis choices work together to determine each experiment's ability to precisely measure absolute IBD detection rates.  
Given the lack of detailed discussion in the literature of absolute reactor and detector systematics for these experiments, we make a variety of educated assumptions about their capabilities in these aspects.

We begin by considering a detector with  $2\times {10}^{29}$ fiducial target protons and 44\% IBD detection efficiency; these values roughly correspond to the size of the PROSPECT detector~\cite{prospect} and the IBD detection efficiency of the PROSPECT and Soli$\delta$ detectors~\cite{prospect,solid}.  
We then consider deployment of this detector at an HEU and LEU reactor site.  For an HEU measurement, we assume deployment of this detector at 7~m distance from a compact reactor core with a thermal power of 85~MW$_{\textrm{th}}$ and a 100\% \uFive~fission fraction; these parameters roughly correspond to the expected parameters of the PROSPECT experiment at the High Flux Isotope Reactor (HFIR) at Oak Ridge National Laboratory, and are also reasonably close to those expected for STEREO and Soli$\delta$.  We also assume a three-year HEU data-taking period that includes six yearly 26-day operating cycles, similar to that experienced by HFIR; these experimental parameters result in 160,000 total IBD detections.  This corresponds to a total signal statistical uncertainty of 0.25\%.  
For an LEU measurement, we consider deployment of the same detector at 20~m distance from a 2.9~GW$_{th}$ reactor with a fission fraction profile matching that of the Daya Bay reactor cores.  This reactor-detector distance roughly corresponds to baseline ranges previously achieved in short-baseline LEU reactor experiments~\cite{bib:danss_2016,bib:B4,rovnomon}.  
These parameters for the LEU case result in $\sim$3500 daily IBD detections.  An 18-month data-taking run spanning an entire LEU fuel cycle is assumed for this study.  
We assume a 1:1 signal-to-background ratio for both LEU and HEU short-baseline experiments.  
Assumed experimental parameters are summarized in Table~\ref{tab:SBLParams}.  

\begin{table}[tb!]
\centering
\begin{tabular}{l|cc}
\hline
\multicolumn{1}{c|}{\textbf{Parameter}} & \multicolumn{1}{c}{\textbf{Value, HEU}} & \multicolumn{1}{c}{\textbf{Value, LEU}} \\ \hline \hline
\textbf{Reactor} && \\ 
 Thermal Power & 85 MW & 2.9 GW \\
 Burnup Profile & - & From~\cite{bib:cpc_reactor} \\ 
 Reactor Up-Time & 47\% & 100\% \\ \hline
\textbf{Detector} && \\ 
Target Protons & \multicolumn{2}{|c}{2$\times 10^{29}$} \\
IBD Detection Efficiency & \multicolumn{2}{|c}{44\%} \\ \hline 
\textbf{Experimental} && \\ 
Core-Detector Distance & 7~m & 20~m \\ 
Data-Taking Length & 3~y & 1.5~y \\ 
Signal-to-Background & 1:1 & 1:1 \\ \hline
\textbf{Uncertainty, Reactor} && \\ 
Thermal Power & 0.5\% & 0.5\% \\
Fission Fractions & - & 0.6\% \\
Energy per Fission & 0.1\% & 0.2\% \\ \hline
\textbf{Uncertainty, Detector} && \\ 
Target Protons & \multicolumn{2}{|c}{1.0\%} \\
Detection Efficiency & \multicolumn{2}{|c}{1.0\%} \\
IBD Cross Section & \multicolumn{2}{|c}{0.1\%} \\ \hline
\textbf{Total Systematics, Uncorrelated} & \textbf{0.5\%} & \textbf{0.8\%} \\ 
\textbf{Total Systematics, Correlated} & \multicolumn{2}{|c}{\textbf{1.4\%}} \\ 
\end{tabular}
\caption{Assumed experimental parameters for the hypothetical future short-baseline reactor experiments described in the text.}
\label{tab:SBLParams}
\end{table}

Systematic uncertainties in these reactor and detector parameters must also be considered.  Thermal power uncertainties of 0.5\% have been achieved in recent LEU experiments~\cite{bib:cpc_reactor,bib:reno,dc_bump}; we assume a similar uncertainty for both HEU and LEU measurements.  
For the LEU measurement, an additional 0.6\% uncertainty~\cite{bib:cpc_reactor,bib:prl_evol} is included to account for a 5\% absolute uncertainty in fission fractions ($\delta F_i/F_i$)~\cite{bib:cpc_reactor}.  
For HEU and LEU cores, uncertainties in the energy released per fission are 0.1\% and 0.2\% respectively~\cite{bib:fr_ma}.  
On the detector side, target proton and detection efficiency uncertainties are both assumed to be 1.0\%; while these are ambitious targets for the next generation of short-baseline reactor experiments, similar values have been reported for previous reactor measurements (For target protons, see Ref.~\cite{bib:cpc_reactor}; for IBD detection, see Refs.~\cite{bib:B4,bib:dc_eff}).  IBD interaction cross-section uncertainties are 0.1\%~\cite{Agashe:2014kda}.  
Total systematic uncertainties for the HEU and LEU cases are thus assumed to be 1.5\% and 1.6\%, respectively.  For a measurement at both HEU and LEU reactors with the same detector, the detector-related uncertainty component, 1.4\%, is assumed to be correlated between measurements.  

For the purpose of simplicity, finite core and detector sizes, variations in IBD rates due to non-equilibrium conditions in the reactors, and IBD contributions from spent or non-fissioning nuclear fuel are not considered; it is expected that corrections for these effects will result in negligible changes to statistical or systematic uncertainties in the IBD yield measurements.  
It is further assumed that any systematic uncertainties due to background estimation and subtraction are also negligible; demonstration of high-precision cosmogenic background reduction for short-baseline reactor experiments utilizing continuous monitoring of fast neutron and muon rates has been previously demonstrated~\cite{bib:nucifer}.  

\subsection{Simulating Hypothetical Future IBD Yield Datasets}

To assess the impact of new IBD yield measurements, we have generated simulated  IBD yield datasets for a variety of experimental scenarios.  
We initially define the underlying physics by assuming in these simulations that the 5.4\% reactor antineutrino flux anomaly arises purely due to improper \uFive~predictions.  
This corresponds to utilizing the previously-cited $\sigma^{th}_{9,8,1}$ values from Refs.~\cite{bib:huber,bib:mueller2011} and a $\sigma_5$ value of 6.04\xSec.  
At a simulated LEU reactor measurement, this procedure will produce results in rough, but not precise, agreement with Daya Bay's existing evolution measurement. 
A simulated HEU measurement will produce results that are not in general agreement with existing HEU measurements.  
This choice, while somewhat arbitrary, enables us to easily examine future experiments' abilities to differentiate between the \uFive-only and common conversion error hypotheses.  
This choice also establishes a cohesive picture between Daya Bay and future hypothetical HEU and LEU measurements.  
The goal of this study is not to unify all existing historical, modern, and future datasets into a common picture, but to provide physics motivation for future short-baseline IBD yield measurements by demonstrating how a progressive set of new yield measurements may improve constraints on $\sigma_i$ values beyond those enabled recently by Daya Bay.

For our study, we utilize five experimental scenarios.  These scenarios are presented in order of experimental precision, from a single Daya Bay-like evolution measurement, to a precise suite of short-baseline HEU and LEU measurements made at different times with the same detector.  

\begin{enumerate}
\item \textbf{Dataset 1:} 
\\\textit{Hypothetical Daya Bay-like LEU measurement.} \\
For this dataset we consider eight data points with fission fractions identical to those reported by Daya Bay~\cite{bib:prl_evol}. 
We also use the quoted Daya Bay statistical and systematic uncertainties (both correlated and uncorrelated) between the data points.  
\item \textbf{Dataset 2:} 
\\\textit{Addition of a short-baseline HEU measurement.} \\
Keeping the assumptions the same for the LEU measurement, we add an HEU measurement. 
For the HEU data point, we utilize the experimental parameters defined in Table~\ref{tab:SBLParams}.  HEU and LEU datasets are assumed to be completely uncorrelated.
\item \textbf{Dataset 3:} 
\\\textit{Improvement in LEU uncertainties.} \\
The uncertainties in the previous LEU measurement are reduced to the values quoted at the bottom of Table~\ref{tab:SBLParams}, while the HEU measurement is left unchanged.  This change corresponds to an improvement of an existing Daya Bay-like LEU experiment.
\item \textbf{Dataset 4:}
\\\textit{Broadening of LEU fission fraction ranges.} \\
The previous LEU measurement is replaced with a new one corresponding to a short-baseline experiment sampling fluxes from a single LEU core, as described in Table~\ref{tab:SBLParams}.  
By sampling a single core, the observed range of fission fractions will increase over that observed in a Daya Bay-like measurement.  
Statistics for this case (2.5 million IBDs) are comparable to that for the previous cases (2.2 million IBDs), with the difference producing negligible impact on the results shown below.  
While detector parameters are identical for the LEU and HEU cases, they are assumed to be separate detectors with uncorrelated uncertainties.  
Such a scenario could be realized through comparison of new and impending short-baseline reactor datasets, such as PROSPECT and DANSS, provided both experiments are capable of reaching the systematics levels listed in Table~\ref{tab:SBLParams}.  
\item \textbf{Dataset 5:} 
\\\textit{Highly-correlated HEU and LEU measurements.} \\
Identical to the previous scenario, but with full correlation between detector systematics in both experiments.  
Such a scenario could be realized through re-deployment of a short-baseline detector at an LEU reactor following completion of an HEU measurement.  
\end{enumerate}


\begin{table*}[t!]
\centering
\begin{tabular}{c|c||c|c|c||c}
\hline
\multirow{2}{*}{Case} & \multirow{2}{*}{Description} & \multicolumn{3}{|c||}{Precision on $\sigma_i$ (\%)} & Common Conversion  \\ \cline{3-5}
& & \uFive &\pNine &\uEight & Error $\overline{\Delta\chi}^2_{min}$ \\ \hline \hline
1& Daya Bay-like LEU & 2.8& 5.9&10.0  & 13.7 \\ \hline
2& Daya Bay-like LEU + new HEU &1.3&5.3 &9.2 & 14.0   \\  \hline
3& Improved Daya Bay-like LEU + HEU  &1.3&4.8&8.9 & 20.2 \\ \hline
4& Short-Baseline LEU + HEU  &1.2 & 3.7&8.8 & 56.2 \\ \hline
5& Short-Baseline LEU + HEU, Correlated  &1.5&3.8 & 6.7 & 56.7  \\ \hline
\end{tabular}
\caption{Constraints on IBD yields of \uFive, \pNine, and \uEight~from various future hypothetical IBD yield datasets.  
Constraints are given as a percentage of the best fit IBD yields.  
Also given for each dataset's $5\times10^6$ simulated experiments are $\overline{\Delta \chi}^{2}_{min}$, the average difference in $\chi^2$ values between the fit described in Eq.~\ref{eq:Iso4} and the more-constrained common conversion error hypothesis fit.  
This metric represents the ability of each dataset to rule out the latter hypothesis, given a true deficit in \uFive~only.  
}
\label{tab:Uncertainty_Improvement}
\end{table*}

\subsection{Impact of Future LEU and HEU IBD Yield Measurements}

For each dataset, we run simulated experiments by varying statistical and systematic factors for each dataset within the 1$\sigma$ bounds described in Table~\ref{tab:SBLParams}.  
For each simulated experiment, we determine best-fit $\sigma_i$ parameters according to Eq.~\ref{eq:Iso4}, which includes 10\% constraints on the two isotopes $\sigma_{8,1}$.  
Without both constraints, the fit is highly degenerate for the Daya Bay-like Dataset 1 defined above.  
Measurement uncertainties on $\sigma_{5,8,1}$ for each dataset are then defined by the 1$\sigma$ range of the full set of simulated experiments.  
We note that because of the two external constraints, a significant improvement in the measurement of \uEight~will be visible as a reduction of $\sigma_8$ measurement precision beneath the external 10\% limit.

The results of the fits to the differing simulated datasets are shown in Table~\ref{tab:Uncertainty_Improvement}. 
As expected, fits to Dataset 1 and Dataset 2 produce $\sigma_{5,1,8}$ measurement precisions nearly identical to the Daya Bay evolution data and the combined fit to all data shown in Figure~\ref{fig:Fit1}.
A systematic improvement in the Daya Bay-like LEU measurement in Dataset 3 produces modest gains in the $\sigma_{9,8}$ values, with the most significant improvement for \pNine.
The transition to a short-baseline LEU measurement sampling a single reactor core in Dataset~4 introduces substantial gains in $\sigma_{9}$, owing to the much larger range of fission fractions accessible to this new detector-core configuration.  
Finally, full correlation between the short-baseline HEU and LEU measurements reduces the $\sigma_{8}$ measurement uncertainty to 6.7\%.  

Final precision on $\sigma_{5,9,8}$ measurements with all considered datasets are comparable or superior to that provided by existing summation- and conversion-based predictions.  
In the case of $\sigma_{8}$ and $\sigma_{5}$, the achieved final uncertainty from \nuebar flux measurements significantly exceeds the claimed 8.15\% and 2.11\% uncertainties utilizing the summation approach and conversion approach, respectively~\cite{bib:mueller2011}.  
For \pNine, the final uncertainty, 3.8\%, is somewhat higher than the 2.45\% uncertainty claimed using the conversion approach.  
We note that the dominant uncertainty in our fit is provided by the inflated 10\% uncertainty in \nuebar production by \pOne~ that was applied to remove degeneracy from the fit with minimal bias towards the Huber-predicted result.  
If we instead \textit{utilize} the Huber-predicted $\sigma_1$ uncertainty of 2.15\%, we obtain a measurement uncertainty of 2.7\% for $\sigma_9$; uncertainty for $\sigma_5$ remains at 1.5\% while the uncertainty for $\sigma_8$ improved slightly to 6.5\%.  
To more robustly address the degeneracy between plutonium isotopes, future IBD yield measurements at reactors containing plutonium-rich mixed-oxide (MOX) fuel may be valuable; many such reactors  are currently in operation across Europe.  
Comparison of IBD yield measurements between reactors utilizing weapons-grade and reactor-grade MOX fuel, as described in Refs.~\cite{bowdenMOX, huberMox}, would be of particular interest for independently constraining \pNine~and \pOne~contributions.

In addition to determining the precision of future isotopic IBD measurements, we can also perform statistical tests on these five simulated datasets to examine various hypotheses regarding the origin of the reactor anomaly.  
Since we have considered hypothetical datasets under the assumption that the reactor flux anomaly arises purely from \uFive, we can easily check how well these simulated datasets disfavor the common conversion error hypothesis, where the anomaly is assumed to arise equally from the predictions of the three beta-converted isotopes.  

We begin by calculating, for each of the five scenarios above, the distribution of $\Delta\chi^2_{min}$ values between the fit in Eq.~\ref{eq:Iso4} and the common conversion error fit in Eq.~\ref{eq:Iso6} for all simulated experiments.  
Average $\overline{\Delta\chi}^2_{min}$ values for each scenario are listed in Table~\ref{tab:Uncertainty_Improvement}.  
The addition of one HEU measurement in Dataset 2 whose result matches the \uFive-only hypothesis will deviate somewhat from the prediction of the common conversion error hypothesis, leading to a small  increase in $\overline{\Delta\chi}^2_{min}$.  
The systematically improved LEU measurement of Dataset 3 then produces a more substantial $\sim$40\% increase in the $\overline{\Delta\chi}^2_{min}$: with increased constraint on the normalization of the LEU data, the fit can not as easily scale all LEU values to match the data's slope to that required by the common conversion error hypothesis.  
As the fission fraction range of the LEU data is increased by sampling a single core with a short-baseline detector in Dataset 4, the $\overline{\chi}^2_{min}$ of the this hypothesis is greatly increased.  
This experimental improvement is clearly the most powerful in terms of enhancing the ability to discriminate between the two highlighted hypotheses.  

\begin{table*}[t!]
\centering
\begin{tabular}{c|c||c|c||c|c||c|c}
\hline
\multirow{3}{*}{Parameter} & \multirow{3}{*}{Value}& \multicolumn{6}{c}{Precision on $\sigma$ (\%) }\\ \cline{3-8}
& & \multicolumn{2}{c}{\uFive}   & \multicolumn{2}{c}{\pNine}& \multicolumn{2}{c}{\uEight} \\ \cline{3-8}
& & D3 & D5 & D3 & D5 & D3 & D5 \\ \hline \hline
None & Default & 1.26 & 1.50 & 4.80 & 3.84 & 8.91 & 6.68 \\ \hline
\multirow{2}{*}{Signal to Background} & 1:2 & 1.27 & 1.51  & 4.80 & 4.15& 8.91 & 6.83  \\  
& 10:1 & 1.25 & 1.49 & 4.80 & 3.40 & 8.91 & 6.53 \\ \hline
\multirow{2}{*}{HEU Reactor Power} & 1.0\% & 1.39 & 1.67 &4.80 & 3.95 & 9.01 & 7.43 \\  
 & 2.0\% & 1.67 & 1.94 & 4.90 & 4.15 & 9.21 & 8.61 \\ \hline 
\multirow{2}{*}{Detector Normalization} & 2.0\% & 1.82 & 2.27 & 5.10 & 4.45 & 9.41& 6.73 \\ 
& 3.0\% & 2.46 & 3.1 & 5.65 & 5.30 & 9.60 & 6.78 \\ \hline
Combined & Worst, Combined & 2.51 & 3.51 & 5.78 & 5.90 &9.68 & 8.71 \\ \hline 
\end{tabular}
\caption{Impact of variations in experimental parameters on future achievable \uFive, \uEight, and \pNine~IBD yield precisions.  Measurement precisions are given as a percentage of the best fit IBD yields for Datasets 3 (D3) and 5 (D5) described above.}
\label{tab:variations}
\end{table*}

For each experimental scenario, we can assign a p-value preference against the common conversion hypothesis for each simulated experiment's $\Delta\chi^2_{min}$ following a methodology similar to that outlined in Ref.~\cite{GiuntiMe} for determining a dataset's preference for one of two non-nested models. 
First, we generate Monte Carlo toys following the common conversion error hypothesis for each of the five experimental scenarios listed above.  
We then produce a distribution of $\Delta \chi^{2}_{min,toy}$ between the less-constrained fit in Eq.~\ref{eq:Iso4} and and the common conversion error fit.  
By comparing a simulated experiment's $\Delta \chi^{2}_{min}$ to this $\Delta \chi^{2}_{min,toy}$ distribution, we obtain a p-value that describes the extent to which the dataset disfavors the common conversion error hypothesis with respect to the less-constrained fit.  
While we find that the $\Delta \chi^{2}_{min,toy}$ distributions for Datasets 1-5 roughly approximate a $\chi^2$ distribution with NDF=1, we provide all quoted numbers  based on the Monte Carlo simulations.  

For Dataset 1, 94\% of all simulated experiments are able to disfavor the common conversion error hypothesis at better than 3$\sigma$ confidence level.  
This result provides an interesting contrast to the Daya Bay fuel evolution dataset, which disfavors the common conversion error hypothesis at slightly less than 3$\sigma$.  
While Daya Bay's data is well-fit by the 235-only hypothesis, it is also more compatible with the common conversion error hypothesis than the 'average' simulated Daya Bay-like 235-only hypothesis dataset.  
This result is not entirely surprising: Daya Bay's best-fit $\sigma_5$ is slightly higher than that needed to support the 235-only hypothesis, while its best-fit $\sigma_9$ is 3\% lower than the Huber-predicted value.  

For Datasets 2, and 3, 95\%, and 97\% of all simulated experiments are able to disfavor the common conversion error hypothesis at better than 3$\sigma$ confidence level, respectively.  
For Datasets 4 and 5 including both LEU and HEU short-baseline yield measurements, 100\% of simulated experiments disfavor the common conversion error hypothesis at better than 5$\sigma$ confidence level.
This result provides a further illustration of the substantial model discrimination capabilities possible with future high-precision HEU and LEU IBD yield measurements.

\subsection{Impact of Variations in Experimental Parameters}
\label{subsec:Syst}

In the previous section we showed that improved isotopic IBD yield constraints are achievable with a set of five progressive measurements with specific experimental parameters described in Table~\ref{tab:SBLParams}.  
However, the ability of upcoming short-baseline experiments to achieve these parameters has yet to be demonstrated; in many cases, their achievement will be a significant challenge.  
To examine the range of possible future constraints on isotopic IBD yields, we examine the impact of variations in key experimental parameters in our analysis.  

Background rejection has been identified as a key consideration for the new generation of short-baseline experiments~\cite{vSBL}, many of which will be deployed with minimal overburden.  
A number of HEU-based experiments have thus far not demonstrated the level of background rejection necessary to achieve the 1:1 signal-to-background ratio (S:B) assumed in this study~\cite{bib:neutrino4,bib:nucifer,bib:danss_2016}.  
On the other hand, other short-baseline efforts have either already achieved a S:B ratio far surpassing 1:1 at LEU reactors \cite{bib:neos}, or have presented prototype- and Monte Carlo-based evidence to support their achievement of a S:B of 1:1 or higher at HEU reactors~\cite{solid,prospect}.  
To examine the impact of S:B on IBD yield precision, we adjust the assumed S:B of all short-baseline measurements in Datasets 2-5 , while continuing to neglect background contributions for the Daya Bay-like measurements of Datasets 1-3.  
Isotopic IBD yield measurement precisions for these variations are shown in Table~\ref{tab:variations} for the 1:2 and 10:1 S:B cases.  
Precisions on $\sigma_5$ and $\sigma_8$, which are dominated by systematic, rather than statistical uncertainties, are only marginally impacted by the assumed S:B.  
The impact is larger for $\sigma_9$, with precision reduced from 3.84\% to 4.15\%.  

For HEU reactors, precision measurement of reactor thermal powers may be more difficult than at LEU reactors.  
To maximize their operational efficiency, commercial LEU facilities generally employ redundant reactor heat flow monitoring systems capable of delivering precise real-time power measurements, as described in Ref.~\cite{bib:cpc_reactor}.  
It is possible that existing online power monitoring systems utilized at research HEU reactor facilities may not be capable of achieving the same level of precision demonstrated at commercial facilities.  
For this reason, we also consider a variation in the assumed thermal power uncertainty of HEU reactors to 1\% and 2\%, while maintaining the default 0.5\% uncertainty for LEU measurements.  
For \uFive~and \uEight, the impact of this uncertainty inflation is significantly larger than the previously-examined S:B variation.  
The greatest impact is on the \uFive~and \uEight~IBD yield measurement, with uncertainties increasing from 1.50\% to 1.94\% and  6.68\% to 8.61\% respectively.  
For \pNine, the reduction in precision is similar in scope to that observed from S:B variations.  

We have also considered variations in achievable detector-related normalization uncertainties.  
Percent-level precision in estimation of timing, pulse-shape, topology, and energy analysis cut efficiencies in a short-baseline detector is likely to be challenging.  
Compared to scintillator experiments where percent-level selection efficiency uncertainties have been achieved, the next generation of short-baseline detectors contain more non-scintillating material and are significantly smaller in size, which are likely to lead to lower efficiencies and larger efficiency variations across a detector.  
For detectors implementing target segmentation and/or fiducialization to aid in background rejection, percent-level target proton uncertainties will also be challenging to achieve.  
Fortunately, a number of upcoming experiments have designs enabling a wide array of detector-internal calibrations, which may enable percent-level precision despite these challenges.  
To examine the impact of detector normalization uncertainties, we have inflated this parameter in our analysis to 2.0\% and 3.0\%.  
This adjustment can be seen as either doubling one of these two sources of uncertainty beyond the default case, or roughly doubling both sources of uncertainty.  
For Dataset 3 and Dataset 4, this change has a more substantial negative impact than either of the two previously-considered parameters.  
In the case of 2\% uncertainty for Dataset 3, \uEight, \pNine, and \uEight~constraints are all worsened, to 1.82\%, 5.10\%, and 9.41\%.  
For Dataset 5, constrains on \uFive~are more significantly worsened, while constraints for \uEight~ are only slightly impacted, due to the correlation between HEU and LEU measurements.  

To understand the required experimental precision needed for gains in IBD yield precision in the future, it is useful to consider a full set of measurements achieving the worst of the experimental parameters considered above.  
In this scenario, the best future short-baseline detector achieves a 3\% detector normalization uncertainty, is operated at an HEU reactor for three years with a 2\% thermal power uncertainty and 1:2 S:B, and then is operated at an LEU reactor for 1 cycle with a 0.5\% thermal power uncertainty and a 1:2 S:B.  
In this scenario, \uFive, \pNine, and \uEight~precisions of 3.51\%, 5.90\%, and 8.71\% are achieved.  
With the exception of \uEight, these constraints do not represent a significant improvement over those currently achieved by the Daya Bay evolution result.  
Thus, if future short-baseline IBD yield measurements wish to provide significant improvement in global knowledge of isotopic IBD yields, they should aim for systematics comparable to those listed in Table~\ref{tab:SBLParams}. 

\section{Summary}  
\label{sec:summary}

Using global fits of IBD yield measurements at reactors of varying fission fractions, we have investigated current and possible future constraints on \uFive, \uEight, and \pNine~\nuebar production in reactors.
We have shown that existing global rate data and Daya Bay's recent fuel evolution results produce differing best-fit values for \uFive~and \pNine~IBD yields.  
These differing values result in differing preferences between datasets for the primary cause of the reactor antineutrino flux anomaly: Daya Bay data favors incorrectly predicted \uFive~IBD yields, while global rate data favors equal responsibility from all beta-converted IBD yield predictions.  
The discontinuity also produces a global fit with an increased $\chi^2_{min}$ relative to fits to the individual datasets, and no significant preference for either reactor anomaly hypothesis.  
We have further shown that the underlying cause of these differences is the discrepancy between Daya Bay and highly-enriched uranium core IBD yield datasets.  

This paper and other recent analyses have suggested a variety of ways in which this discontinuity in the global IBD yield picture can be alleviated.  
First, the issue could obviously be resolved by assuming flaws in either the HEU or Daya Bay datasets, and discarding that dataset from consideration.  
On the other hand, the discontinuity could be reduced while still including all datasets.  
Ref.~\cite{GiuntiMe} has shown that models including both sterile neutrinos and incorrect flux predictions provide a good fit to global IBD yield data.  
In addition, we have shown in this analysis that an improved combined fit is obtained by allowing incorrect \uEight~predictions in addition to incorrect \uFive~and \pNine~predictions.  
If the latter scenario is used and fit to all data, we obtain the first direct isotopic IBD yield measurement for \uEight, 7.02 $\pm$ 1.65 \xSec, a value nearly 2$\sigma$ below existing predictions.  

By applying global fits to future hypothetical IBD yield datasets, we have demonstrated how a future experimental short-baseline reactor program can distinguish between the various hypotheses described above and address all primary aspects of the reactor flux anomaly.  
We first note that short-baseline reactor experiments must directly address the sterile neutrino hypothesis using methods insensitive to underlying flux models~\cite{prospect,solid,bib:danss_2016,stereo}; this essential measurement, which can be done at both HEU and LEU reactors, removes fit degeneracies between sterile oscillations and incorrect flux predictions.  
Next, short-baseline experiments with even modest precision at HEU reactors will be valuable in providing a cross-check of historical HEU IBD yield measurements.  
Given the existing inconsistencies between historical HEU measurements, such as those in Ref.~\cite{bib:srp} previously noted in Ref.~\cite{bib:chao}, such cross-checks seem necessary to increase confidence in global fit-derived isotopic IBD yields.  
Similarly, cross-checks would also be valuable at LEU reactors -- in particular, an independent validation of Daya Bay's new evolution analysis by a different experiment.  
Upcoming short-baseline experiments can serve both of the purposes outlined above.  

If future short-baseline experiments are capable of achieving IBD yield measurements with $\sim$1.5\%-level precision, they can produce excellent constraints on individual isotopic IBD yields.  
Naturally, as previously demonstrated in~\cite{Giunti}, high-precision HEU measurements can help solidly constrain \uFive~yields.  
Due to their single-core sampling and increased range of samples of fission fractions, short-baseline LEU experiments, when combined with a new HEU measurement, can significantly constrain \pNine~yields.   
To maximally constrain the \uEight~isotopic IBD yield, measurements should be made at both HEU and LEU reactors with successive deployments of a single detector.  
Achievement of the full experimental program outlined above would produce isotopic IBD yield measurements of \uFive, \pNine, and \uEight~with precisions of 1.5\%, 3.8\%, and 6.7\%.  
Naturally, as constraints on individual isotopes are improved, so is the ability to identify the individual isotopes responsible for the reactor antineutrino flux anomaly.  
Combined HEU and LEU IBD yield measurements made with successive deployments of a single detector would be capable of distinguishing at better than 5$\sigma$ confidence level whether conversion-based IBD yield predictions are incorrect for only one isotope or are equally incorrect for all isotopes.  

In addition to addressing the sterile neutrino hypothesis as it relates to reactor neutrinos, the described future experimental program would provide \uFive, \pNine, and \uEight~IBD yield constraints superior to those achievable by any existing reactor flux prediction.  
The resulting precision neutrino-based flux model would enable low-uncertainty flux predictions for future experiments at any reactor type, or at any isolated point in the fuel cycle of an LEU reactor.  
Such a model could benefit analyses attempting to measure particle physics parameters via comparisons between differing reactor cores;  
One such example is unitarity tests of the PMNS matrix~\cite{unitarity}, which utilize comparisons between $\theta_{13}$ experiments and medium-baseline reactor experiments, such as JUNO~\cite{juno3}.
Improved models would also benefit future reactor-based precision tests of the Standard Model utilizing alternate interaction channels, such as $\overline{\nu}_e$-e scattering and coherent neutrino scattering~\cite{connie,gemma}.  
Precision neutrino-based IBD yield predictions can also be compared to existing summation predictions to diagnose possible problems in existing community-standard nuclear databases.

\section{Acknowledgements}
This work was supported by DOE Office of Science, under award No. DE-SC0008347, as well as by the IIT College of Science.  We thank Carlo Giunti, Tom Langford and David Jaffe for their useful comments and discussion.

\bibliographystyle{apsrev4-1}
\bibliography{References}{}

\end{document}